\documentclass{article}

\PassOptionsToPackage{numbers, compress}{natbib}
\usepackage[preprint]{neurips_2023}

\usepackage[utf8]{inputenc} 
\usepackage[T1]{fontenc}    
\usepackage{hyperref}       
\usepackage{url}            
\usepackage{booktabs}       
\usepackage{amsfonts}       
\usepackage{nicefrac}       
\usepackage{microtype}      
\usepackage{xcolor}         
\usepackage{amsmath}
\usepackage{amssymb}
\usepackage{mathtools}
\usepackage{amsthm}
\usepackage{multirow}
\usepackage{wrapfig}
\usepackage[capitalize,noabbrev]{cleveref}
\usepackage{graphicx}
\usepackage{subfigure} 
\usepackage{float}
\usepackage{algorithm} 
\usepackage{algorithmic}
\usepackage{textcomp} 
\usepackage{listings} 
\usepackage{enumitem}
\usepackage{utfsym}
\usepackage{times}
\usepackage{latexsym}
\usepackage{fancyvrb}
\usepackage{CJKutf8}

\newcommand{\commentout}[1]{}

\title{TouchASP: Elastic Automatic Speech Perception that Everyone Can Touch}

\author{
  Technical Report \thanks{
    Authors: Xingchen Song\textsuperscript{1,2}, Chengdong Liang\textsuperscript{1,2}, Binbin Zhang\textsuperscript{1,2}, Pengshen Zhang\textsuperscript{1}, ZiYu Wang\textsuperscript{1}, Youcheng Ma\textsuperscript{1}, Menglong Xu\textsuperscript{1,2}, Lin Wang\textsuperscript{1}, Di Wu\textsuperscript{1,2}, Fuping Pan\textsuperscript{2}, Dinghao Zhou\textsuperscript{2}, Zhendong Peng\textsuperscript{2}.
    Corresponding author: Xingchen Song (sxc19@tsinghua.org.cn)
  } \\
  GUA Speech Team\textsuperscript{1}, \quad WeNet Community\textsuperscript{2}
}

\begin{document}

\maketitle

\begin{abstract}
    \label{sec:abs}
    
    Large Automatic Speech Recognition (ASR) models demand a vast number of parameters, 
    copious amounts of data, and significant computational resources during the training process. 
    However, such models can merely be deployed on high-compute cloud platforms and are only capable of performing speech recognition tasks. 
    This leads to high costs and restricted capabilities.
    In this report, we initially propose the elastic mixture of the expert (eMoE) model. 
    This model can be trained just once and then be elastically scaled in accordance with deployment requirements. 
    Secondly, we devise an unsupervised data creation and validation procedure and gather millions of hours of audio data from diverse domains for training.
    Using these two techniques, our system achieves elastic deployment capabilities while reducing the Character Error Rate (CER) on the SpeechIO testsets from 4.98\% to 2.45\%.
    Thirdly, our model is not only competent in Mandarin speech recognition but also proficient in multilingual, multi-dialect, emotion, gender, and sound event perception.
    We refer to this as Automatic Speech Perception (ASP), and the perception results are presented in the experimental section.

    \end{abstract}

\section{Introduction}
\label{sec:intro}

\begin{figure}[!ht]
    \centering
    \includegraphics[width=1.0\textwidth]{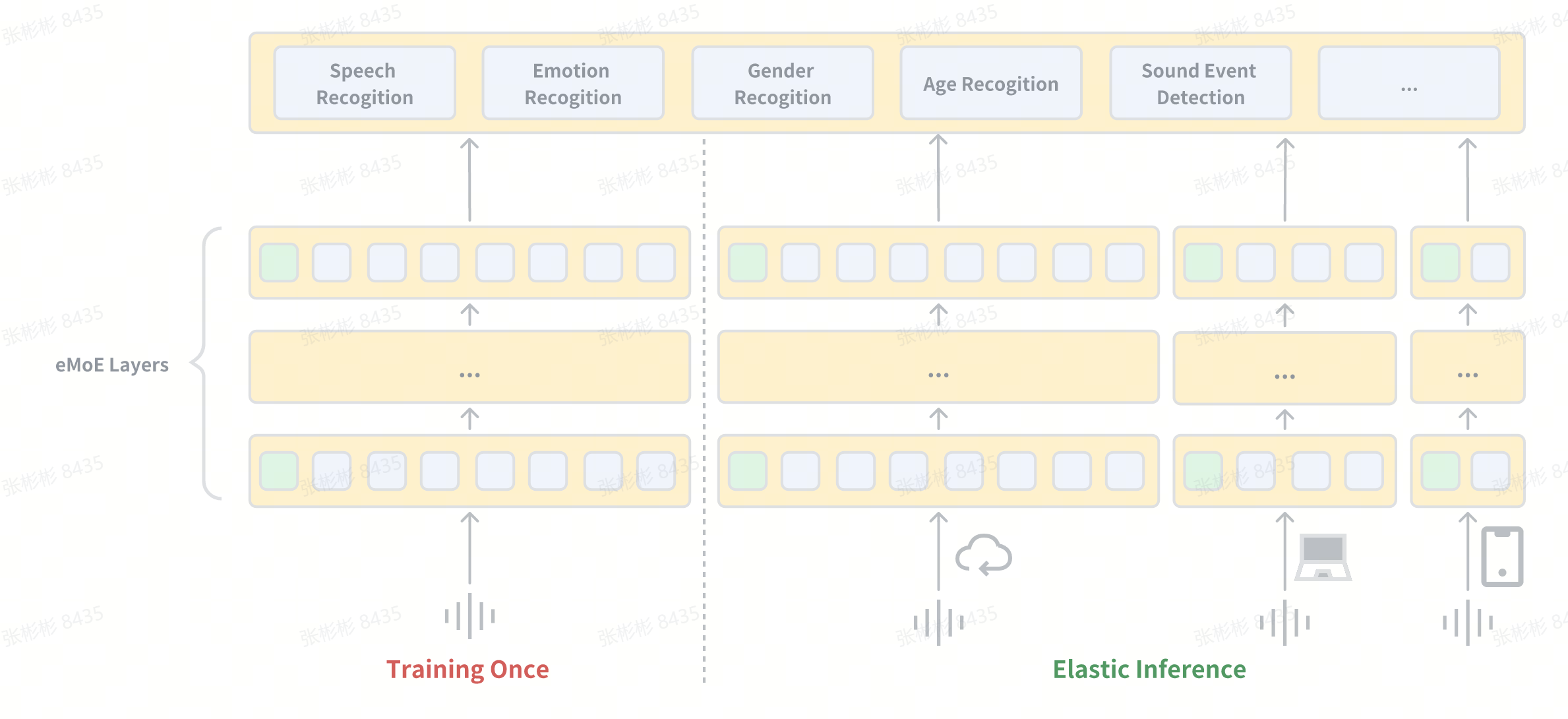}
    \caption{TouchASP: Train once, Elastic inference with multiple speech perception capabilities}
    \label{fig:touchasp}
\end{figure}

In recent years, large models have achieved remarkable progress, and their astonishing capabilities have been validated across a multitude of fields, with the domain of speech perception being no exception. However, while model scaling can bring better capabilities, the training cost is admittedly high, which makes it less accessible to everyone.

Automatic speech recognition, as an important part of Artificial General Intelligence (AGI), has a great deal of outstanding work that has verified the effectiveness of model scaling.  \cite{shazeer2017outrageously,lepikhin2020gshard,jiang2024mixtral,dai2024deepseekmoe,yang2024qwen2,pan2024dense,you2021speechmoe,hu2023mixture,wang2023language,song2024u2++}. 
Nevertheless, in real business scenarios, when the product needs to be switched to different deployed devices, especially those with different capabilities, more training costs and a longer product cycle will be required.
In order to make sure that more people benefit from large models, we propose the elastic mixture of the expert (eMoE) model based on DeepSeek \cite{dai2024deepseekmoe} in Section \ref{sec:emoe}. 
During training, the eMoE employs the dynamic expert strategy, and during inference, the number of experts is pruned to adapt to devices with different capabilities. This can significantly save training costs while maintaining high recognition accuracy.

Another important factor that limits large model from being accessed is the amount of data. Therefore, we devise a simple, reproducible and high-retention data processing pipeline in Section \ref{sec:datapipeline}. With this process, we have collected 1,000k hours of paired speech data in diverse domains and styles from the Internet. More importantly, our experimental results show that the collected data has a positive incentive for Character Error Rate (CER). Similar conclusions have also been verified in TouchTTS \cite{song2024touchtts}.

As is known to all, the larger the model is, the greater its capabilities will be. In order to make better use of the capabilities of large model and the potential of massive amounts of data, inspired by Whisper \cite{radford2023robust}, we introduce Automatic Speech Perception (ASP), which is a multi-task perception framework, in Section \ref{sec:gasp}. It extends the eMoE model from ASR task to more speech perception tasks. Experiments show that it not only performs excellently in multilingual, multi-dialect recognition tasks but also achieves good results in language identification, age identification, gender identification, and emotion identification tasks.

Figure \ref{fig:touchasp} shows the framework of our system, we named it TouchASP, 
which can be trained once, and inferenced elastically with multiple speech perception capabilities.

\section{Related Work}

Model scaling is a crucial technique in large models and Mixture of Experts(MoE)\cite{shazeer2017outrageously,lepikhin2020gshard}  is one of most significant techniques for model scaling. MoE empowers models to be pretrained with considerably less compute. This means that you can scale up the model or dataset size to a much greater extent with the same compute budget as a dense model. Specifically, a MoE model should attain the same quality as its dense counterpart much more rapidly during pre-training. So far, there are many typical works about MoE in LLM fundation model, such as Mixtral\cite{jiang2024mixtral}, DeepSeekMoE\cite{dai2024deepseekmoe}, QWen\cite{yang2024qwen2}, etc. In particular, DeepseekMoE introduces two principal strategies
of fine-grained expert segmentation and shared expert isolation. DS-MoE \cite{pan2024dense} achieves strong computation and parameter efficiency by employing dense
computation across all experts during training and sparse computation during inference. 
In Speech, SpeechMoE \cite{you2021speechmoe} explores the MoE based model for speech recognition.
\cite{hu2023mixture} proposes a streaming truly multilingual Conformer incorporating MoE.
LR-MoE\cite{wang2023language} extracts language-specific representations through MoE, which is guided to learn by a framewise language routing mechanism. U2++ MOE\cite{song2024u2++} simplify the integration
of MoE and preclude the necessity for any auxiliary losses for streaming speech recognition.

Data scaling is of fundamental significance in the learning of large models, where trillions of tokens are typically employed for the training of large language models\cite{dubey2024llama}. In speech area, GigaSpeech \cite{chen2021gigaspeech} introduces a multi-domain English
speech recognition corpus with 10,000 hours of high quality labeled
audio suitable for supervised training, and 40,000 hours of total audio
suitable for semi-supervised and unsupervised training. 
WenetSpeech \cite{zhang2022wenetspeech} and WenetSpeech4TTS \cite{ma2024wenetspeech4tts} release a 10000+ hours multi-domain mandarin corpus for speech recognition and
a 12,800-hour mandarin corpus for Text to Speech(TTS) respectively.
Emilia \cite{he2024emilia} introduces an extensive, multilingual, and diverse speech dataset for large-scale speech generation with over 101k
hours of speech across 6 languages.
Whisper \cite{radford2023robust} scales to 680,000 hours of multilingual and multitask supervision, then Whisper-3 scales 1 million hours of weakly labeled audio and 4 million hours of pseudo-labeled audio collected using Whisper-2. 
USM \cite{zhang2023google} applys pre-training the encoder of the model on a large unlabeled multilingual dataset
of 12 million hours spanning over 300 languages, and fine-tuning on a smaller labeled dataset. 
MMS \cite{pratap2024scaling} scales to 1 million hours data with 1000+ languages.
Seed-ASR \cite{bai2024seed} is training on over 20 million hours of speech data and nearly 900 thousand hours of paired ASR data.
SenseVoice \cite{an2024funaudiollm} is trained with over 40k hours of data, supporting more than 50 languages.

Capability scaling is one of the greatest advantages of large models, as we can train just one model on multiple tasks, and the learning of these multiple tasks can benefit each other. 
By training on a multi-task format, Whisper\cite{radford2023robust} could perform transcription, translation, voice activity detection, alignment, and language identification.
MMS \cite{pratap2024scaling} built pre-trained wav2vec 2.0 models covering 1,406 languages, a single multilingual automatic speech recognition model for 1,107 languages, speech synthesis models for the same number of languages, as well as a language identification model for 4,017 languages.
SenseVoice \cite{an2024funaudiollm} is designed as a speech foundation model with multiple speech understanding capabilities, including automatic speech recognition, spoken language identification, speech emotion recognition, and audio event detection.
QWen-Audio \cite{chu2023qwen} scales up audio-language pre-training to cover over 30 tasks and various audio types, such as human speech, natural sounds, music, and songs, to facilitate universal audio understanding abilities.
Llama-3 \cite{dubey2024llama} and Gemini \cite{team2023gemini} and other LLM based model demonstrate great capabilities and scalability in speech content understanding.

\section{Methods}

\subsection{eMoE: Elastic Mixture of Expert}
\label{sec:emoe}

\begin{figure}[!ht]
    \centering
    \includegraphics[scale=0.16]{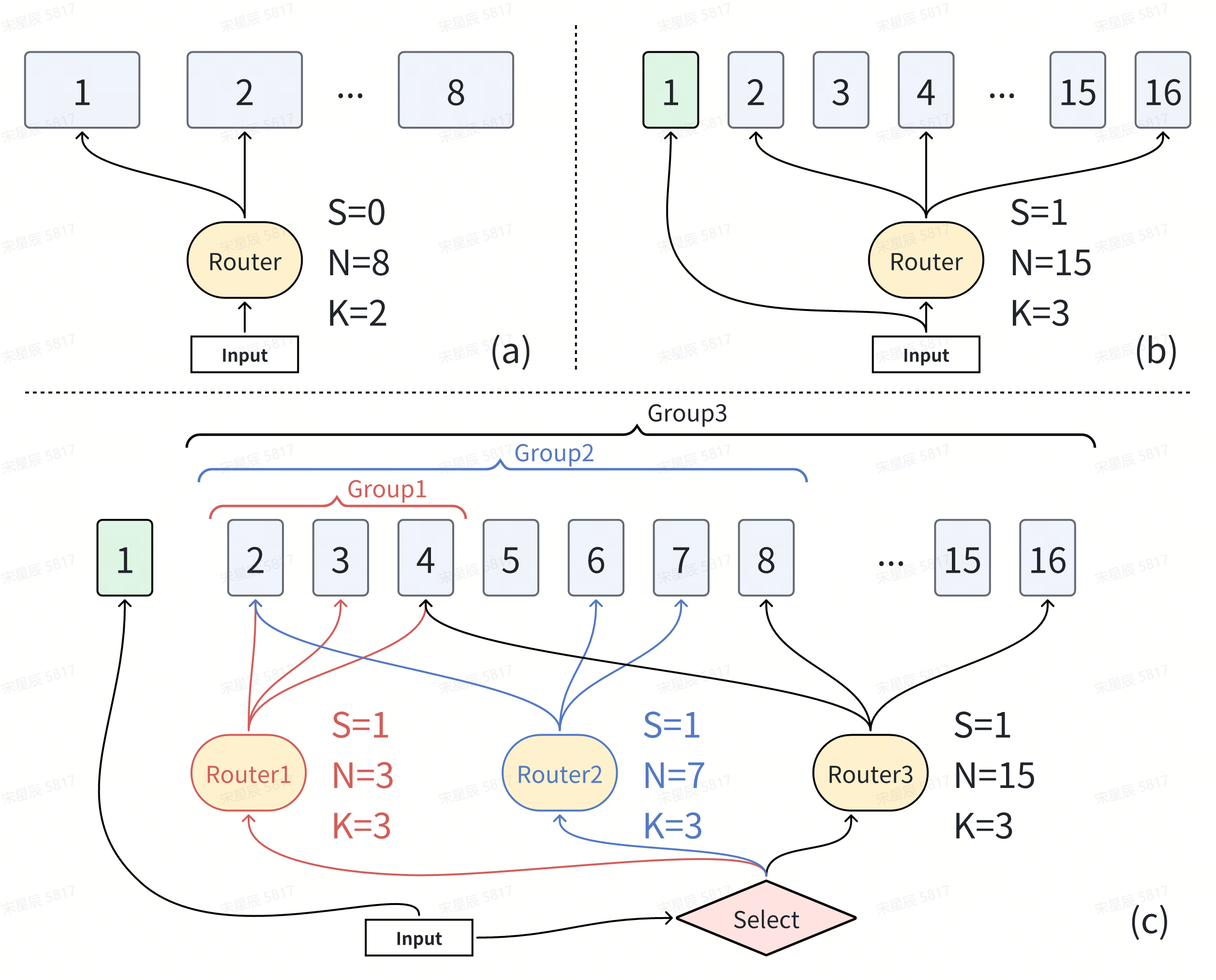}
    \caption{Illustration of Elastic MoE (eMoE). 
    Subfigure (a) showcases the conventional top-K routing ($K=2$) among N experts ($N=8$). 
    Subfigure (b) illustrates the fine-grained expert segmentation and the shared expert isolation strategy that has been originally proposed in DeepSeekMoE, 
    with one shared expert ($S=1$) as an illustrative example. 
    Subsequently, subfigure (c) demonstrates the integration of DeepSeekMoE and our dynamic training approach, constituting the complete eMoE architecture. 
    It is noteworthy that across these three architectures, the number of expert parameters and computational costs remain constant.
    }
    \label{fig:emoe}
\end{figure}


Motivated by slimmable networks\cite{yu2018slimmable}, the core idea of eMoE is to use a dynamic expert strategy during training, so that during inference,
experts can be elastically pruned based on runtime resource constraints. 
which permits instant and adaptive accuracy-efficiency trade-offs at runtime.
Moreover, since MoE is structured and organized by experts, it naturally enables dynamic training and pruning of MoE at the expert level without any additional work.


As shown in Figure \ref{fig:emoe}, in terms of model architecture, eMoE adopts a similar architecture to DeepSeekMoE, 
which uses shared experts and a larger number of experts to achieve better model performance.
In eMoE, all experts at each layer are organized into multiple groups.
Assuming there are $M$ experts in total, including $S$ shared and $N$ independent experts (as shown in the figure, $M=S+N$),
the $i$-th expert is denoted as $e_i$, and the smallest group contains $G$ experts (in the figure, $G=4$), with $G>=K$, where $K$ represents the number of experts to be routed.
The experts contained in group 1, group 2, group 3 and group Z are respectively:
$$g_1=(e_0, e_1, ..., e_G)$$
$$g_2 = (e_0, e_1, ..., e_{2G})$$
$$g_3 = (e_0, e_1, ..., e_{4G})$$
$$...$$
$$g_Z = (e_0, e_1, ..., e_M)$$
where the number of experts in group 2 is twice that of group 1, the number of experts in group 3 is twice that of group 2,
and so on, with a total of $Z$ groups, where $2^{Z-1}*G=M$.

During training, we dynamically and randomly select a group $g_i$ from $(g_1, g_2, ..., g_Z)$,
and only use the experts contained in $g_i$ for training, that is, only select the top-K from the experts contained in $g_i$ for training.
Through the dynamic expert training strategy, our model can learn to dynamically use any group of expert configurations from $(g_1, g_2, ..., g_Z)$ for inference.
If uniform distribution sampling is used, since the later groups contain the experts of the previous groups,
the experts in the previous groups will get more training times than the experts in the later groups, leading to unbalanced training.
To solve this problem, we first use independent and different Routers for different groups, that is, different groups use independent router parameters.
Then, we set different sampling probabilities for different groups, giving smaller probabilities to the previous groups,
and larger probabilities to the later groups, so that the later groups can be selected with a higher probability to alleviate the problem of unbalanced training.
As shown in Figure \ref{fig:emoe}, red, blue, and black respectively show the selection of $K$ experts from different groups $g_1$, $g_2$, $g_3$ for training,
and use their respective independent router.


For the MoE layer, the overall parameter size of the model is related to the number of experts it contains. Therefore, the parameter size of group 1 is twice that of group 0, group 2 is twice that of group 1, and so on.
During inference, we can use only the parameters contained in group $g_0$ for inference, which has the fewest parameters and can be deployed on platforms with limited memory and computing power, such as edge devices.
Alternatively, we can use all the parameters contained in the full group $g_Z$ for inference, which has the largest parameter size and the best performance, suitable for deployment on high-performance platforms.
You can also select the appropriate parameters from $(g_0, g_1, ..., g_Z)$ for elastic deployment based on your platform's resource conditions.

The experts grouping strategy used in this paper is a power strategy, and the experiments are also designed according to the power strategy.
However, we believe that eMoE is also suitable for linear configuration, and you can use a linear strategy for training and inference according to your needs.

\subsection{Data Pipeline}
\label{sec:datapipeline}

\begin{figure}[htbp]
    \centering
    \includegraphics[width=1.0\textwidth]{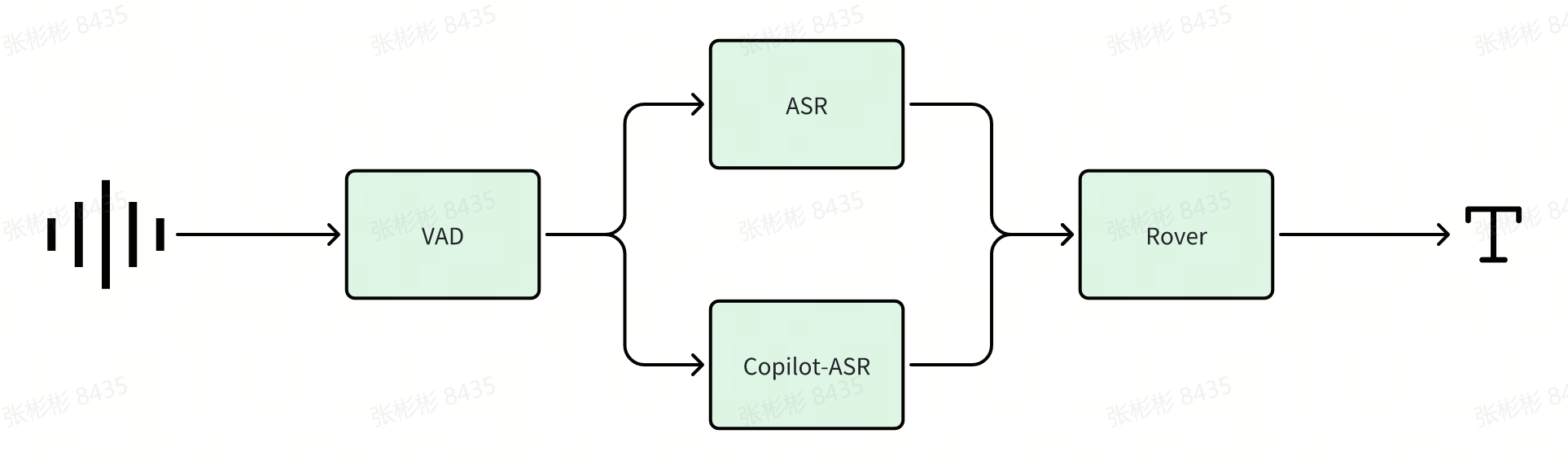}
    \caption{Overview of Our Data Pipeline}
    \label{fig:pipeline}
\end{figure}


Following the trend of recent work\cite{zhang2022wenetspeech, radford2023robust, zhang2023google, pratap2024scaling} 
leveraging web data from the internet for
training speech recognition learning systems, we take a minimalist approach to data pre-processing.
We obtain data from various fields of videos and audiobooks from the internet, including TV dramas, interviews, commentaries, documentaries, technology, games, sports, arts, cuisine, etc.
This results in a very diverse dataset covering a broad distribution of audio from many 
different recording setups, speakers, environments, and styles.
This diversity in audio and fields helps our model train better and more robustly.


To obtain the corresponding labels for the web audio,
As shown in Figure \ref{fig:pipeline}, we propose a simple data processing pipeline that can convert a large amount of raw speech data into ASR training data.
The pipeline mainly consists of the following four modules: \textit{VAD}, \textit{ASR}, \textit{Copilot-ASR}, and \textit{Rover}.
First, we use \textit{VAD} to segment the audio into clips ranging from 2s to 30s, then use the \textit{ASR} model to convert the audio into text.
Next, we use the \textit{Copilot-ASR} model for secondary transcription, and finally use the \textit{Rover} module to filter out data with large discrepancies based on the consistency of the two transcriptions, obtaining the final annotation results.
We use whisperX\cite{bain2022whisperx} as the \textit{VAD} and \textit{ASR} engine, and Paraformer\cite{gao2022paraformer} as the \textit{Copilot-ASR} engine.
We implement the \textit{Rover} function by comparing the WER (Word Error Rate) and PER (Phone Error Rate) between the two transcriptions.
We filter out data with WER greater than 10 and PER greater than 5.
We found that a single ASR model may make mistakes, but the probability of both ASR models making mistakes simultaneously is significantly lower.
Usually, overly low-quality data will show great differences on different ASR models,
so we filter out low-quality data by comparing the consistency of the two transcriptions, thereby improving data quality.



In the end, we processed 650,000 hours of data usable for training from 1.26 million hours of raw data, achieving an overall data retention rate of 51.6\%.
It is worth noting that even with the same data processing pipeline, data from different fields will have significant differences in retention rates.
For example, audiobook data has a relatively fixed recording environment and less background noise,
so the data retention rate will be higher than that of outdoor live broadcast data. The 51.6\% is the overall average after combining data from all fields.

In addition, we collected 150,000 hours of data from open-source data (mostly ASR data),
plus some internal data (also mostly ASR data), forming our one million-hour training dataset.
\subsection{General Perception}
\label{sec:gasp}

Although the recognition of text content within a given audio clip is the core of the speech recognition problem and has been extensively researched, it is not the sole element of the speech perception system. A fully functional speech perception system may comprise numerous additional components, such as speech activity detection, speaker diarization, emotion perception, and non-linguistic audio event detection. These components are typically processed independently, resulting in a relatively complex speech perception system.
To alleviate this complexity, we hope to have a single model that can perform the entire speech perception process rather than just the core speech recognition portion. Therefore, we need to consider how to design the model's input and output so that it can execute different tasks.

To achieve this objective, inspired by Whisper \cite{radford2023robust} and Qwen-Audio \cite{chu2023qwen}, we design a multi-task learning framework so that a single model can handle multiple tasks, eliminating the cumbersome switching when processing different tasks. More significantly, tasks can mutually benefit from each other during joint training. First, similar tasks can gain from knowledge sharing and collaborative learning since they both focus on the basic information embedded in the audio signal. Second, tasks that rely on lower-level perceptual abilities can assist tasks that require higher-level understanding or reasoning. 

\begin{figure}[!ht]
    \centering
    \includegraphics[scale=0.08]{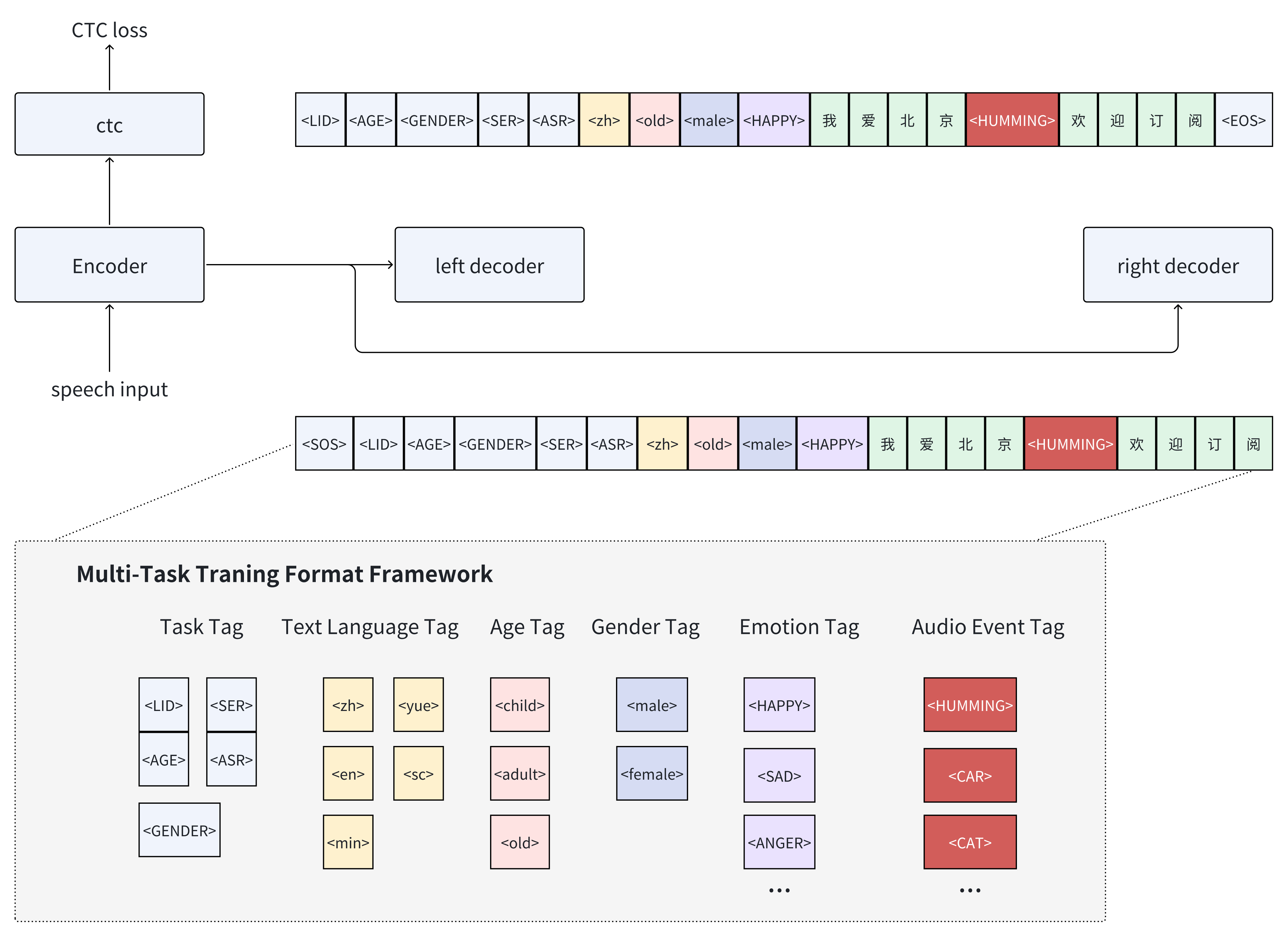}
    \caption{Overview of model structure and multi-task training. The encoder-decoder model is trained on many different speech tasks. All these tasks are represented by a token sequence predicted by the decoder, allowing a single model to replace different stages in the traditional speech processing pipeline. The multi-task training format uses a set of special tokens as task instructions or classification targets, as explained in Section \ref{sec:gasp}.
    }
    \label{fig:gasp}
\end{figure}

The multi-task training framework is depicted in Figure \ref{fig:gasp}. All tasks are represented by a token sequence predicted by the decoder, as detailed below:

\begin{itemize}
\item Task Tag: The subsequent tokens specify the particular task. We categorize audio tasks into five categories, namely language identification, age identification, gender identification, emotion identification, and speech transcription.
\item Text Language Tag: We support five language identification tasks, including Mandarin, English, Cantonese, Sichuan dialect, and Minnan dialect.
\item Age Tag: We support three age identification tasks, namely child, adult, and elderly.
\item Gender Tag: We support two gender identification tasks, namely male and female.
\item Emotion Tag: We support seven emotion identification tasks, namely neutral, anger, sadness, happiness, surprise, fear, and disgust.
\item Audio Event Tag: We support detecting over seventy common sound events. For instance, cat meows, dog barks, phone calls, humming, keyboard clicks, and other sounds commonly encountered in human-machine interaction scenarios. We define sound event detection as a detection task so that it can be predicted simultaneously in CTC and the decoder.
\end{itemize}

TouchASP employs the eMoE (Section \ref{sec:emoe}) structure, supports using CTC to output text transcription and sound event labels, and supports using the decoder to rescore the CTC results. The left-decoder is utilized for self-regressive decoding, and the results of language detection, age detection, gender detection, emotion detection, sound event detection, and text transcription are output. In conclusion, TouchASP is a simple yet flexible model that can handle multiple speech tasks and can select different decoding methods according to actual requirements.

\section{Experiments}

We obtained a total of 1 million hours of data through the data pipeline proposed in this paper combined with internal data.
In terms of experimental design, we first verify the effectiveness of our data pipeline through data scaling experiments by testing speech recognition rates with different data volumes.
Then, we verify the effectiveness of the eMoE algorithm on 1M hours of data.
Finally, based on the above results, we add multi-task capabilities to the 1 million-hour eMoE model to verify its general perception ability.


We use the SpeechIO\footnote{https://github.com/SpeechColab/Leaderboard} test set to evaluate the performance of our data crawling pipeline and the eMoE algorithm.
SpeechIO is a large, robust, comprehensive benchmarking test set for Mandarin Automatic Speech Recognition.
SpeechIO test sets are carefully curated by SpeechIO authors, 
crawled from publicly available sources (Youtube, TV programs, Podcast etc), 
covering various well-known scenarios and topics, transcribed by paid professional annotators.
We use 26 publicly available test sets from SpeechIO, including independent test sets such as news, interviews, sports, speeches, education, games, live broadcasts, audiobooks, commentaries, and cross talks,
with a total of 60.2 hours, averaging 2.3 hours per test set.
We use the weighted average WER on these 26 test sets as the final evaluation metric.

All of our experiments are carried out within the WeNet \cite{yao2021wenet} toolkit. 
We make use of 80-dimensional log-mel filterbank features, 
which are computed by means of a 25ms window that shifts every 10ms. 
Global mean and variance normalization is applied to each frame. 
In terms of modeling unit, character based representations are adopted for Mandarin,
and byte pair encoding (BPE) is utilized for English.

\subsection{Data Pipelines Evaluation}


To verify the correctness of the data pipelines, we trained for 100K steps on 160K hours, 500K hours, and 1M hours of data using the U2++ MoE\cite{song2024u2++} model architecture.
The final performance was evaluated using WER. The details of the data are shown in Table \ref{tab:data}.

\begin{table}[]
    \caption{Details of the 160K/500K/1M hours data}
    \label{tab:data}
    \centering
    \begin{tabular}{p{0.1\textwidth}p{0.8\textwidth}}
    \toprule
    Dataset & Description                                                                                             \\ \midrule
    160K & 160K hours of high-quality annotated internal data                                                  \\ \midrule
    500K & 220K hours of data generated by the proposed pipeline, 160K hours of internal data, and additional open-source data                                 \\ \midrule
    1M   & 650K hours of data generated by the proposed pipeline, 160K hours of internal data, and additional open-source data \\ \bottomrule
    \end{tabular}
\end{table}

\begin{table}[ht]
    \centering
    \caption{SpeechIO CER on 160K/500K/1M hours data}
    \label{tab:data_scale}
    \begin{tabular}{llll}
    \toprule
    Data & 160K & 500K & 1M   \\ \midrule
    CER & 3.8  & 3.44 & \textbf{3.09} \\ \bottomrule
    \end{tabular}
\end{table}


As shown in Table \ref{tab:data_scale}, as we add more data generated by the proposed pipeline,
we achieve better CER on the SpeechIO test set. Ultimately, with the data reaching 1 million hours, we achieve significant gains.
This directly demonstrates the effectiveness of the proposed pipeline and once again verifies the importance of data scale for speech recognition.

\begin{figure}[htbp]
    \centering
    \includegraphics[width=0.9\textwidth]{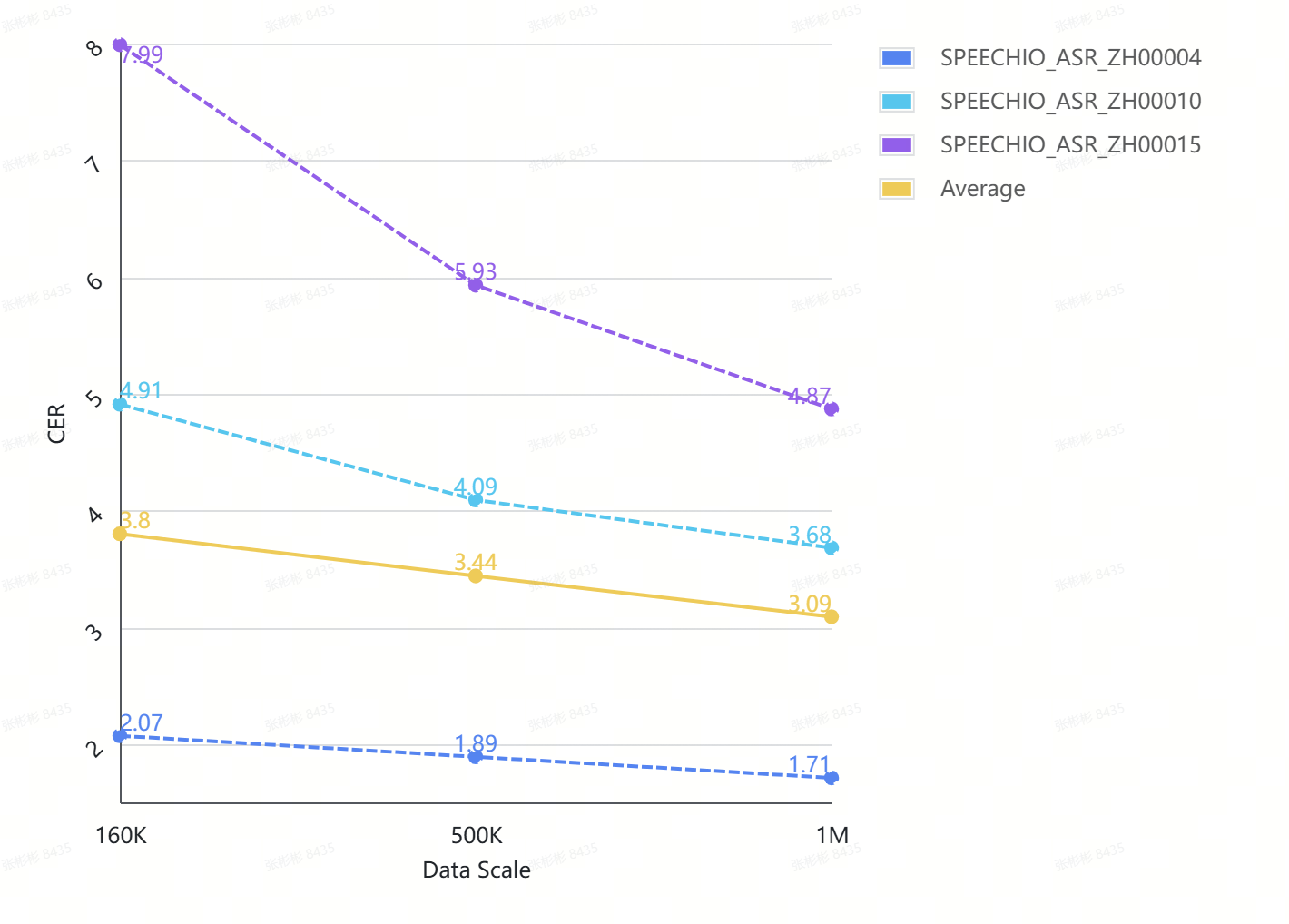}
    \caption{Typical test sets CER on SpeechIO when data scales. \textit{SPEECHIO\_ASR\_ZH00004} is a simple talk testset,
    \textit{SPEECHIO\_ASR\_ZH00010} is a medium difficulty interview testset,
    and \textit{SPEECHIO\_ASR\_ZH00015} is a difficult story imitation testset.}
    \label{fig:scale}
\end{figure}

To further investigate the impact of data scale on different scenarios and test sets,
we selected three typical test sets of varying difficulty and scenarios from the 26 test sets in SpeechIO.
Figure \ref{fig:scale} shows the changes in CER for each test set as the data scales.
As can be seen, when the data scale increases, we achieve consistent improvements across the three test sets.
In fact, we achieve consistent improvements on the vast majority of test sets in SpeechIO.
\subsection{eMoE Evaluation}

In this section, we aim to answer three questions:
\begin{enumerate}
    \item Does the DeepSeek-MoE architecture perform better than the regular MoE in speech recognition tasks, and if so, what is the optimal model configuration?
    \item Can eMoE work effectively?
    \item Using 1M hours of data and eMoE, what performance can we ultimately achieve compared to our previous work and others' work?
\end{enumerate}

\begin{table}[ht]
    \caption{Best eMoE configuration exploration}
    \label{tab:moe_config}
    \centering
    \begin{tabular}{@{}lc|ccc@{}}
    \toprule
    Model         & MoE-1B          & \multicolumn{3}{c}{DeepseekMoE-1B}                \\ \midrule
    MoE config    & S=0, N = 8, K=2 & S=1, N=15, K=3 & S=2, N=30, K=6 & S=4, N=60, K=12 \\
    Training Time & 2 days            & 4 days           & 5.5 days         & 8.5 days          \\
    CER  & 4.44            & 3.62           & 3.6            & 3.54            \\ \bottomrule
    \end{tabular}
\end{table}

Let's first look at the first question. We trained approximately 50k steps using both the regular MoE architecture and the DeepSeekMoE architecture based on 500K data.
As shown in Table \ref{tab:moe_config}, where S represents shared experts, N represents non-shared experts, and K represents the top-K selected for routing.
First, comparing MoE and DeepSeekMoE, we can see that DeepSeek-MoE achieves significant improvements compared to MoE, at the cost of doubling the training time.
Then, for DeepSeekMoE, we trained with three configurations, using increasingly refined and more experts for training.
It can be seen that more experts can bring better recognition results, but the training time will also increase significantly.
To balance performance and training time, we use the configuration \textit{S=1, N=15, K=3} for the subsequent experiments.

\begin{table}[ht]
    \caption{eMoE vs DeepSeekMoE}
    \label{tab:emoe}
    \centering
    \begin{tabular}{@{}c|c|c|c|c|c@{}}
    \toprule
                     & DeepSeekMoE-350M & DeepSeekMoE-1B & \multicolumn{3}{c}{eMoE-1B}                           \\ \midrule
    Training  & S=1, N=3, K=3    & S=1, N=15, K=3 & \multicolumn{3}{c}{S=1, N=sample{[}3, 7, 15{]} ,K=3} \\
    Infernece & N=3              & N=15           & N=3              & N=7              & N=15            \\
    CER     & 3.99             & 3.62           & 3.95             & 3.71             & 3.68            \\ \bottomrule
    \end{tabular}
\end{table}

Then, let's answer the second question. Using the same 500K data, we designed two different sizes of DeepSeekMoE architectures as baselines.
\textit{DeepSeekMoE-350M} is used as the lower bound for our comparison,
\textit{DeepSeekMoE-1B} is used as the upper bound for our comparison,
both models are trained using regular training methods.
For eMoE, we use the previously mentioned method for dynamic training during training, and use 3, 7, and 15 experts for inference respectively.
When $N = 3$, the inference parameter size of eMoE is consistent with DeepSeekMoE-350M.
When $N = 15$, the inference parameter size of eMoE is consistent with DeepSeekMoE-1B.
It can be seen that under the same inference parameter size, eMoE, which is trained only once, can achieve comparable results to multiple different DeepSeekMoE models.

\begin{table}[]
\caption{Comparison to our previous work and others' work}
\label{tab:emoe_comparison}
\centering
\begin{tabular}{@{}cccc@{}}
\toprule
Model                 & Data                & Model Size & SpeechIO CER \\ \midrule
\multicolumn{4}{c}{Commercial API SpeechIO Leaderboard\footnote{https://github.com/SpeechColab/Leaderboard}}   \\ \midrule
Aliyun API            & NA                  & NA         & 1.80          \\
Microsoft API         & NA                  & NA         & 1.95         \\
iFlytek API           & NA                  & NA         & 3.02         \\
Tencent API           & NA                  & NA         & 3.20          \\ \midrule
\multicolumn{4}{c}{Ours}                                                \\ \midrule
Dense                 & 160K                & 225M       & 4.98         \\
MoE                   & 160K                & 1B         & 3.15         \\
\multirow{2}{*}{eMoE} & \multirow{2}{*}{1M} & 335M(N=3)  & 2.67         \\
                      &                     & 1B(N=15)   & \textbf{2.45}         \\ \bottomrule 
\end{tabular}
\end{table}


Based on the answers to the above two questions, we further used 1M hours of data and continued training for 650K steps using the eMoE architecture and training strategy.
Then we obtained Table \ref{tab:emoe_comparison}, where we also listed the results of many commercial APIs on the SpeechIO Leaderboard.
As can be seen, compared to our previous work, using 1M hours of data and eMoE, we optimized the recognition rate on SpeechIO from 4.98 to 2.49,
achieving a leap forward.
Compared with other commercial APIs, we also achieved competitive results.
\subsection{General Perception Evaluation}

For the evaluation of general perceptual ability, we employed a one-million-hour eMOE model as the base model. Subsequently, we conducted training by incorporating dialect ASR data, age data, gender data, emotion data, and sound event data. 
The data details are shown in Table \ref{tab:multitask_data}.

\begin{table}[ht]
\caption{Multi-task training datasets.}
\label{tab:multitask_data}
\centering
\begin{tabular}{cccc p{0.5\linewidth}}
\toprule
Type                    & Task                                                               & Language                                        & Hour                                                            & Description                                                                                                                                                                                                                                                                                   \\ \midrule
\multirow{18}{*}{Speech} & \multirow{6}{*}{\begin{tabular}[c]{@{}c@{}}ASR\\ LID\end{tabular}} & Yue                                             & $\sim$5k                                                        & The data consists of open-source data such as Common Voice \cite{ardila2019common} and mdcc \cite{yu2022automatic}, as well as other manually annotated data Cantonese data.                                                                                                               \\
                        &                                                                    & Sichuan                                         & $\sim$3k                                                        & The data employs manually annotated Sichuan data, encompassing dialogue and reading scenarios.                                                                                                                                                                                                \\
                        &                                                                    & MinNan                                          & $\sim$1.7k                                                      & The data utilizes open-source data Minspeech \cite{lin2024minspeech}.                                                                                                                                                                                                            \\ \noalign{\vspace{2pt}} 
                        \cline{2-5} 
                        \noalign{\vspace{2pt}} 
                        & \multirow{6}{*}{\begin{tabular}[c]{@{}c@{}}Gender\\ Age\end{tabular}}               & \multirow{6}{*}{\begin{tabular}[c]{@{}c@{}}ZH\\ EN\end{tabular}} & \multirow{6}{*}{\begin{tabular}[c]{@{}c@{}}2k (gender)\\ 100 (age)\end{tabular}} & The data employs open-source data sources including Aishell \cite{bu2017aishell}, Librispeech \cite{panayotov2015librispeech}, Kespeech \cite{tang2021kespeech}, and kaggle data, as well as wake-up data. Kaggle data and wake-up data are both annotated with age and gender labels, whereas the other data sources only have gender labels.                             \\ 
                        \noalign{\vspace{2pt}} 
                        \cline{2-5} 
                        \noalign{\vspace{2pt}} 
                        & \multirow{4}{*}{SER}                                                                & \multirow{4}{*}{\begin{tabular}[c]{@{}c@{}}ZH\\ EN\end{tabular}} & \multirow{4}{*}{$\sim$100}                                                       & The data employs open-source datasets namely M3ed \cite{zhao2022m3ed} , meld \cite{poria2018meld} , iemocap \cite{busso2008iemocap} , esd \cite{keshtiari2015recognizing} , ravdess \cite{livingstone2018ryerson} , and tts-db30, along with other manually annotated data. \\ \midrule
\multirow{4}{*}{Sound}                   & \multirow{4}{*}{AED}                                                                & \multirow{4}{*}{N/A}                                             & \multirow{4}{*}{$\sim$3k}                                                        & The data uses open-source data ESC-50 \cite{piczak2015esc} and FSD50K \cite{fonseca2021fsd50k} as sound event data. Then, the data is expanded through random splicing with ASR data.                                                                                        \\ \bottomrule
\end{tabular}
\end{table}

\textbf{Multilingual Speech Recognition}.
We utilize Word Error Rate (WER) to evaluate the recognition ability of the model in five languages, namely Mandarin, English, Cantonese, Minnan, and Sichuan. Prior to computing WER, we employ text normalization to standardize the actual transcription data and model prediction data. All Chinese characters are converted to simplified Chinese, and English is converted to uppercase.
The results presented in the Table \ref{tab:dialect_res} demonstrate the comparison of Whisper, SenseVoice, Seed-ASR\cite{bai2024seed}, eMoe-1B, and TouchASP on popular open speech recognition benchmark datasets, including WenetSpeech, Librispeech, Common Voice, and Minspeech. For the Sichuan dialect, manually annotated ASR data is utilized. It can be observed that TouchASP significantly outperforms its counterparts, in most test sets with the exception of Librispeech.
Especially in the dialect data, TouchASP achieved the best performance. 
The performance of the Minnan dialect is relatively weak. This may be attributed to insufficient data availability. Morever, recognition tasks for the Minnan dialect are more challenging.
Compared with eMoe-1B, TouchASP exhibits a substantial decline in Mandarin performance. This indicates that incorporating dialect data can have an impact on Mandarin recognition.

\begin{table}[ht]
\centering
\caption{Performance comparison of different models within the corpora of Mandarin, English, Cantonese, Sichuan dialect, and Minnan dialect using the WER metric.}
\label{tab:dialect_res}
\begin{tabular}{cccccc}
\toprule
dataset                   & Whisper-L-V3  & SenseVoice-L & Seed-ASR & eMoE-1B       & \multicolumn{1}{l}{TouchASP} \\ \midrule
wenetspeech test\_net     & 10.48         & 6.01         & \textbf{4.66}     & 5.32 & 5.52               \\
wenetspeech test\_meeting & 18.87         & 6.73         & 5.69     & \textbf{5.42} & 5.94               \\
Librispeech test\_clean   & \textbf{1.82} & 2.57         & 1.58     & 2.03          & 2.03                         \\
Librispeech test\_other   & 3.50 & 4.28         & \textbf{2.84}     & 4.28          & 4.38                         \\
CommonVoice yue           & 10.41         & 6.78         & -        & -             & \textbf{4.15}                         \\
mdcc yue                  & 14.6          & -            & -        & \textbf{-}    & \textbf{2.85}                \\
sichuan diagule           & -             & -            & -        & \textbf{-}    & \textbf{12.93}               \\
sichuan reading           & -             & -            & -        & -             & \textbf{5.00}                         \\
Minspeech test            & -             & -            & -        & -             & \textbf{28.11}                        \\ \bottomrule
\end{tabular}

\end{table}

\textbf{Language Identification}: 
We utilize the F1-score to evaluate the language identification capability. TouchASP can classify Mandarin, English, Cantonese, Minnan dialect, and Sichuan dialect. The test data employs the same data as ASR, with Wenetspeech test set being replaced by Aishell test data. This is due to the fact that Wenetspeech data contains English and data with dialectal accents.
The results are presented in Table \ref{tab:lid}. We have found that TouchASP is an excellent language identification model, surpassing Whisper. In the Mandarin, English, and Cantonese data sets, the performance of TouchASP is close to that of SenseVoice. Morever, TouchASP has a good classification ability for Sichuan and Minnan dialects.

\begin{table}[ht]
\centering
\caption{Comparison of F1-Score between TouchASP, Whisper, and Sensevoice in the language identification task.}
\label{tab:lid}
\begin{tabular}{llllll}
\toprule
                                                           & zh       & en       & yue      & sichuan   & min       \\
\midrule
Whisper-L-V3                                            & 0.487 & 0.423 & 0        & -         & -         \\
SenseVoice-S & \textbf{0.995}    & \textbf{0.994}   & \textbf{0.998} & -         & -         \\
TouchASP                                                   & 0.990 & 0.988 & 0.995  & \textbf{0.995} & \textbf{0.996} \\
\bottomrule
\end{tabular}
\end{table}

\textbf{Gender and Age Recognition}:
As indicated in Table \ref{tab:general_result}, the model demonstrates excellent recognition capability for gender and age. In comparison with Qwen-Audio, the accuracy rate of gender classification is enhanced to 99\%, while the accuracy rate of age classification is increased to 78\%. The reason for the relatively low accuracy of age recognition is due to insufficient age data. In addition, age recognition is more challenging.

\begin{table}[ht]
\centering
\caption{The results of Gender Recognition, Age Recognition, Speech Emotion Recognition (SER) and Sound Event Detection (SED). The gender test set consists of Aishell1, kws, and kaggle data. The age data consists of kws and kaggle data. The detailed description of the data is shown in Table~\ref{tab:multitask_data}.}
\label{tab:general_result}
\begin{tabular}{llll}
\toprule
Task                    & test set                                    & model        & ACC (\%)   \\
\midrule
\multirow{2}{*}{Gender} & \multirow{2}{*}{gender test} & Qwen-Audio   & 61.0  \\
                        &                                            & TouchASP     & \textbf{99.2}  \\
\midrule
\multirow{2}{*}{Age}    & \multirow{2}{*}{age test}          & Qwen-Audio   & 28.0   \\
                        &                                            & TouchASP     & \textbf{80.0}  \\
\midrule
\multirow{4}{*}{SER}    & \multirow{4}{*}{meld}                      & EmoBox \cite{ma2024emobox}       & 51.9  \\
                        &                                            & Sensevoice-S \cite{an2024funaudiollm} & 57.8  \\
                        &                                            & Sensevoice-L \cite{an2024funaudiollm} & \textbf{63.1}  \\
                        &                                            & TouchASP     & 50.5  \\
\midrule
\multirow{3}{*}{SED}    & \multirow{3}{*}{esc-50}                    & PANNs \cite{kong2020panns}        & 83.3  \\
                        &                                            & MSM-MAE (with SSL) \cite{niizumi2022masked}                   & 85.6  \\
                        &                                            & SS-AST (with SSL) \cite{gong2022ssast}                    & 88.8  \\
                        &                                            & BEATS (with SSL) \cite{chen2022beats}        & \textbf{95.6}  \\
                        &                                            & TouchASP     & 85.7 \\
\bottomrule
\end{tabular}
\end{table}

\textbf{Speech Emotion Recognition}: We evaluate speech emotion recognition on the meld dataset. We report classification accuracy and compare them with some recently published SER benchmark(EmoBox) \cite{ma2024emobox}. As indicated in Table \ref{tab:general_result}, it is evident that TouchASP has demonstrated good performance in speech emotion recognition. Nevertheless, when compared with SenseVoice, its performance is relatively inferior. The possible reason for this might be the insufficient quantity of speech emotion data, which makes it difficult for the model to learn the characteristics of speech emotion. In the future, we will expand the amount of speech emotion data to enhance speech emotion recognition ability.

\textbf{Audio Event Detection}:  TouchASP is capable of recognizing over 70 types of sound events that may occur during human-machine interaction, including coughing, sneezing, breathing, crying, and the sound of a cat. We conduct a comparison between TouchASP and the state-of-the-art audio event detection models, namely PANN \cite{kong2020panns}, MSM-MAE \cite{niizumi2022masked}, SS-AST \cite{gong2022ssast}, and BEATS \cite{chen2022beats}, based on the ESC-50 dataset. As shown in Table \ref{tab:general_result}, it is found that TouchASP demonstrates good sound event detection performance. However, compared to self supervised learning (SSL) models, its performance is slightly inferior, which may be due to the SSL model using a large amount of data during pre training.



\section{Limitations and Future Works}


In this work, we still use the classic encoder-decoder architecture and predefined tasks to achieve multilingual, multi-dialect, and multi-perception capabilities. 
This framework's scalability is much worse compared to pure LLMs, and it cannot achieve more generalized speech content understanding.
In the future, we will explore large speech models based on LLMs that are multilingual, multi-dialect, multi-perception, and have understanding capabilities.
Furthermore, combined with our TouchTTS \cite{song2024touchtts} work, we hope to create an end-to-end speech interaction model and architecture that can understand and generate speech content simultaneously.
Perhaps we will call it TouchChat, stay tuned!

\bibliographystyle{unsrt}
\bibliography{refs}

\end{document}